\documentclass[aps,prd,superscriptaddress]{revtex4}
\usepackage{epsfig,epsf}
\usepackage{amsmath}
\usepackage{amsthm}
\usepackage{amsfonts}
\usepackage{amssymb}
\usepackage{dsfont}
\usepackage{multirow}
\usepackage{appendix}
\usepackage{slashed}
\usepackage[active]{srcltx}
\usepackage{psfrag}

\begin{document}

\title{Interpretation of the newly discovered $\Omega$(2012)}
\date{\today}
\author{T.~M.~Aliev}
\affiliation{Physics Department,
Middle East Technical University, 06531 Ankara, Turkey}
\author{K.~Azizi}
\affiliation{Physics Department, Do\u gu\c s University,
Ac{\i}badem-Kad{\i}k\"oy, 34722 Istanbul, Turkey}
\affiliation{School of Physics, Institute for Research in Fundamental Sciences (IPM), P. O. Box 19395-5531, Tehran, Iran}
\author{Y.~Sarac}
\affiliation{Electrical and Electronics Engineering Department,
Atilim University, 06836 Ankara, Turkey}
\author{H.~Sundu}
\affiliation{Department of Physics, Kocaeli University, 41380 Izmit, Turkey}

\begin{abstract}
Very recently  Belle Collaboration reported observation of a narrow state called $\Omega(2012)$ with mass $2012.4 \pm 0.7$~(stat)~$\pm 0.6$(syst)~MeV and width $6.4^{+2.5}_{-2.0}$~(stat)~$\pm 1.6$~(syst)~MeV. We calculate the mass and residue of $\Omega(2012)$ state by employing the QCD sum rule method.  Comparison of the obtained results with the experimental data  allows us to interpret this state as $1P$ orbital excitation of the ground state $\Omega$ baryon, i.e. with quantum numbers $J^P=\frac{3}{2}^-$.

\end{abstract}

\maketitle

The theoretical calculations of different parameters of hadrons and  comparison of the obtained results with existing experimental data  not only test our present knowledge on these states but also provide insights in the search for new states. Despite the fact that, in the hadronic sector, many particles were observed so far and their properties were intensively  studied there are still much work to do. Even for the hadrons containing only the light quarks,  their excited states  require more investigations. The quark model  predicts some baryonic excited states that have not yet been observed in the experiment. Searching for these missing baryon resonances attracts attention of not only the experimentalists but also the theoreticians. To understand and identify such states it is necessary to broaden the studies on these baryons.

As a result of these circumstances, the recent observation of the Belle Collaboration has attracted the attentions. They reported  observation of $\Omega(2012)$ with mass  $2012.4 \pm 0.7$~(stat)~$\pm 0.6$(syst)~MeV and width $6.4^{+2.5}_{-2.0}$~(stat)~$\pm 1.6$~(syst)~MeV~\cite{Yelton:2018mag} with a conclusion that it has more likely a spin-parity $J^P=3/2^-$. To date, there are a few $\Omega$ baryons listed in the Particle Data Group (PDG)~\cite{Patrignani:2016xqp}. Only one of them, which is the ground state $\Omega(1672)$, is well established and we have lack of certain knowledge on the nature of the others. To identify the spectrum of the $\Omega$ states, the first orbital excitation of $\Omega(1672)$ state was investigated with different models. The quark model~\cite{Chao:1980em,Kalman:1982ut,Capstick:1986bm,Loring:2001ky,Liu:2007yi,Pervin:2007wa,An:2013zoa,An:2014lga,Faustov:2015eba}, lattice gauge theory~\cite{Engel:2013ig,Liang:2015bxr} and Skyrme model~\cite{Oh:2007cr} are among those studies whose predictions gave consistent mass values with the experimental result of the Belle Collaboration. This may be taken as a support for $\Omega(2012)$ being an orbital excitation of $\Omega(1672)$. To identify the properties of the $\Omega(2012)$ baryon it would also be helpful to investigate its other properties besides the mass. Its strong decay was studied recently in~\cite{Xiao:2018pwe} using chiral quark model and as a result, the possibility of  $\Omega(2012)$ being a $J^P=3/2^-$ was underlined, but it was also stated that the results obtained for the possibility of its being a $J^P=1/2^-$ or $J^P=3/2^+$ are also consistent with the results of the Belle Collaboration within the uncertainties. There are also some studies on the radiative decays~\cite{Kaxiras:1985zv} and magnetic moments of negative parity baryons~\cite{Aliev:2014pfa,Aliev:2015}.

The Refs.~\cite{Chao:1980em,Aliev:2016jnp} present the predictions on the mass of  radially excited decuplet baryons. The prediction of ~\cite{Chao:1980em} for $2S$ state is 2065~MeV and it is close to the mass of $\Omega(2012)$. On the other hand the prediction given by Ref.\cite{Aliev:2016jnp} is $2176\pm 219$~MeV obtained using QCD sum rule method. If we consider the central value of this result, it is larger than the observed mass of $\Omega(2012)$. Therefore in order to get new information about the identification of the nature of $\Omega(2012)$ it is necessary to investigate the mass of the orbital excitation of $\Omega(1672)$, which we represent as excited $\Omega$ state in the remaining part of the discussion. Taking this motivation in hand in the present study, we make a QCD sum rule calculation for the mass of the excited $\Omega$ state. The QCD sum rule method~\cite{Shifman:1978bx,Shifman:1978by,Ioffe81} is among the powerful nonperturbative methods used in the literature extensively with considerable success. To make the calculations in this method one follows three steps. First one is the calculation of a given correlation function in terms of hadronic degrees of freedom (hadronic side). The next step is the calculation of the same correlator in terms of QCD  degrees of freedom (theoretical or QCD side). And final step comprises of the match of the results of these previous two steps considering the coefficients of the same Lorentz structure from both sides.

For the present calculation the mentioned correlation function is as follows:
\begin{equation}
\Pi _{\mu \nu}(q)=i\int d^{4}xe^{iq\cdot x}\langle 0|\mathcal{T}\{J_{\mu
}(x){J^{\dagger}}_{\nu }(0)\}|0\rangle ,  \label{eq:CorrF1}
\end{equation}
Here $J_{\mu}$ is the interpolating current for the state of interest written in terms of quark fields. In the calculations we will use two forms of the interpolating current: 
\begin{eqnarray}\label{Eq:Current}
J_{\mu(+)}&=& \epsilon^{abc} (s^{aT}C\gamma_\mu s^{b})s^{c} ,
\end{eqnarray}
\begin{eqnarray}\label{Eq:Current2}
J_{\mu(-)}&=& \epsilon^{abc} (s^{aT}C\gamma_\mu s^{b})\gamma_5 s^{c},
\end{eqnarray}
where $a$, $b$ and $c$ represent the color indices and $C$ is the charge conjugation operator. The subscript $(+)$ and $(-)$ denote the parities of the corresponding interpolating currents. The main peculiarity of these currents is that they interact with both parities.

The hadronic representation of the correlator is obtained by inserting a complete set of hadronic states in the correlator given in Eq.~(\ref{eq:CorrF1}). For positive parity current, by isolating the ground and first orbital excitation, we get
\begin{eqnarray}
\Pi_{\mu\nu(+)}^{\mathrm{Had}}(q)&=&\frac{\langle 0|J_{\mu(+) } |+(q,s)\rangle \langle +(q,s)|J^\dagger_{\nu(+)}|0\rangle}{q^{2}-m_{+}^{2}}
+\frac{\langle 0|J_{\mu(+) } |-(q,s)\rangle \langle -(q,s)|J^\dagger_{\nu(+)}|0\rangle}{q^{2}-m_{-}^{2}}
+\ldots,
\label{eq:phys}
\end{eqnarray}
where the $|+(q,s)\rangle $ and $|-(q,s)\rangle $ represent the ground state  $\Omega(1672)$  and its first orbital excitation $\Omega(2012)$, respectively  and $m_{+}$ and $m_{-}$ are the corresponding masses. The dots are used to represent the contributions coming from higher states and continuum. The matrix elements between vacuum and one particle states present in Eq.~(\ref{eq:phys}) are parameterized as
\begin{eqnarray}
\langle 0|J_{\mu(+) }|+(q,s)\rangle &=&\lambda_{+}u_{\mu}(q,s),
\nonumber \\
\langle 0|J_{\mu (+)} |-(q,s)\rangle
 &=&\lambda_{-}\gamma_5 u_{\mu}(q,s),
\label{eq:Res}
\end{eqnarray}
where $ \lambda_{+}(\lambda_{-}) $ stands for the residue of the corresponding baryon. Performing the summation over spins of spin-$\frac{3}{2}$ baryons with the help of the formula 
\begin{eqnarray}\label{Rarita}
\sum_s  u_{\mu} (q,s)  \bar{u}_{\nu} (q,s) &= &-(\!\not\!{q} + m_{B})\Big[g_{\mu\nu} -\frac{1}{3} \gamma_{\mu} \gamma_{\nu} - \frac{2q_{\mu}q_{\nu}}{3m^{2}_{B}} +\frac{q_{\mu}\gamma_{\nu}-q_{\nu}\gamma_{\mu}}{3m_{B}} \Big],
\end{eqnarray}
the result of the physical part takes the form
\begin{eqnarray}\label{PhyssSide}
\Pi_{\mu\nu(+)}^{\mathrm{Had}}(q)&=&-\frac{\lambda_{+}^{2}}{q^{2}-m_{+}^{2}}(\!\not\!{q} + m_{+})\Big[g_{\mu\nu} -\frac{1}{3} \gamma_{\mu} \gamma_{\nu} - \frac{2q_{\mu}q_{\nu}}{3m^{2}_{+}} +\frac{q_{\mu}\gamma_{\nu}-q_{\nu}\gamma_{\mu}}{3m_{+}} \Big]\nonumber \\
&-&\frac{\lambda_{-}^{2}}{q^{2}-m_{-}^{2}}(\!\not\!{q} - m_{-})\Big[g_{\mu\nu} -\frac{1}{3} \gamma_{\mu} \gamma_{\nu} - \frac{2q_{\mu}q_{\nu}}{3m^{2}_{-}} +\frac{q_{\mu}\gamma_{\nu}-q_{\nu}\gamma_{\mu}}{3m_{-}} \Big]+\ldots.
\end{eqnarray}
It should be noted that the current $J_{\mu(+)}$ couples not only to spin-3/2 baryons but also to spin-1/2 states. To remove the contribution of unwanted states having spin-1/2 we will choose the proper Lorentz structure which is free from the spin-1/2 pollution. The contribution of the spin-1/2 states is determined by the matrix element
\begin{eqnarray}
\langle 0|J_{\mu}|\frac{1}{2}(q)\rangle =A(\gamma_{\mu}-\frac{4q_{\mu}}{m_{\frac{1}{2}}})u(q).
\end{eqnarray}
From here we see that the terms proportional to $\gamma_{\mu}$ or $q_{\mu}$ contain spin-1/2 contributions. To avoid this pollution we chose the structures $ \!\not\!{q}g_{\mu\nu}$  and $ g_{\mu\nu} $ which solely contain contributions coming from spin-3/2 states. With this consideration the result becomes
\begin{eqnarray}
\Pi _{\mu \nu(+)}^{\mathrm{Had}}(q)&=&-\frac{\lambda_{+}^2}{q^{2}-m_{+}^{2}} \left( \!\not\!{q}g_{\mu\nu}+m_{+}g_{\mu\nu} \right)
-
\frac{\lambda_{-}^2}{q^{2}-m_{-}^{2}} \left( \!\not\!{q}g_{\mu\nu}-m_{-}g_{\mu\nu}\right) +\ldots.
\label{eq:CorFun1}
\end{eqnarray}

The hadronic side of the correlation function for the second current given in Eq.~(\ref{Eq:Current2}) can be obtained from Eq.~(\ref{eq:CorFun1}) with the following replacements: $\lambda_{+}\rightarrow \lambda_{-}^{\prime}$, $\lambda_{-}\rightarrow \lambda_{+}^{\prime}$, $m_{+} \rightarrow m_{-}$ and $m_{-} \rightarrow m_{+}$. Applying the Borel transformation with respect to $(-q^{2})$ in order to suppress the contributions coming from higher states and continuum finally we get the following results for the hadronic sides
\begin{eqnarray}
\mathcal{\widehat B}\Pi _{\mu \nu(+)}^{\mathrm{Had}}(q)&=&\lambda_{+}^2 e^{-\frac{m_{+}^{2}}{M^{2}}} \left( \!\not\!{q}g_{\mu\nu}+m_{+}g_{\mu\nu} \right)
+
\lambda_{-}^2 e^{-\frac{m_{-}^{2}}{M^{2}}} \left( \!\not\!{q}g_{\mu\nu}-m_{-}g_{\mu\nu}\right) +\cdots,\nonumber\\
\label{eq:CorFunBorel}
\end{eqnarray}
\begin{eqnarray}
\mathcal{\widehat B}\Pi_{\mu \nu(-)}^{\mathrm{Had}}(q)&=&\lambda_{-}^{\prime 2} e^{-\frac{m_{-}^{2}}{M^{2}}} \left( \!\not\!{q}g_{\mu\nu}+m_{-}g_{\mu\nu} \right)
+
\lambda_{+}^{\prime 2} e^{-\frac{m_{+}^{2}}{M^{2}}} \left( \!\not\!{q}g_{\mu\nu}-m_{+}g_{\mu\nu}\right) +\cdots.\nonumber\\
\label{eq:CorFunBorel}
\end{eqnarray}
In the next part of discussion we denote the coefficient of the Lorentz structure $ \!\not\!{q}g_{\mu\nu}$ as $\Pi_{1}$ and that of Lorentz structure $g_{\mu\nu}$ as $\Pi_{2}$, correspondingly.

After completing the calculations in the hadronic side now we turn our attention to calculate the correlation function from QCD side using operator product expansion. As an example we present its expression by using the interpolating current given in  Eq.~(\ref{Eq:Current}). The calculation leads to the result
\begin{eqnarray}\label{corre4}
\Pi_{\mu\nu(+)}^{\mathrm{OPE}}(q) &=& \epsilon_{abc}\epsilon_{a'b'c'}\int d^4 x e^{iqx} \langle 0 |\left\lbrace  S^{ca'}_{s}(x)\gamma_{\nu}\widetilde{S}^{ab'}_{s}(x)\gamma_{\mu}S^{bc'}_{s}(x)\right. -S^{ca'}_{s}(x)\gamma_{\nu}\widetilde{S}^{bb'}_{s}(x)\gamma_{\mu}S^{ac'}_{s}(x)\nonumber\\
&-&S^{cb'}_{s}(x)\gamma_{\nu}\widetilde{S}^{aa'}_{s}(x)\gamma_{\mu}S^{bc'}_{s}(x)+S^{cb'}_{s}(x)\gamma_{\nu}\widetilde{S}^{ba'}_{s}(x)\gamma_{\mu}S^{ac'}_{s}(x)-S^{cc'}_{s}(x)Tr\left[ S^{ba'}_{s}(x)\gamma_{\nu}\widetilde{S}^{ab'}_{s}(x)\gamma_{\mu}\right] \nonumber\\
&+&\left. S^{cc'}_{s}(x)Tr\left[ S^{bb'}_{s}(x)\gamma_{\nu}\widetilde{S}^{aa'}_{s}(x)\gamma_{\mu}\right] \right\rbrace  |0\rangle ,
\end{eqnarray}
with $ \widetilde{S}(x)=CS^{T}(x)C $  and the $S_{s}^{ab}(x)$ is the light quark  propagator which is given as
\begin{eqnarray}
 S_{q}^{ab}(x)&=&i\frac{x\!\!\!/}{2\pi^{2}x^{4}}\delta_{ab}-\frac{m_{q}}{4\pi^{2}x^{2}}\delta_{ab}-\frac{\langle
 \overline{q}q\rangle}{12}\Big(1-i\frac{m_{q}}{4}x\!\!\!/\Big)\delta_{ab}-\frac{x^{2}}{192}m_{0}^{2}\langle
 \overline{q}q\rangle\Big( 1-i\frac{m_{q}}{6}x\!\!\!/\Big)\delta_{ab}-\frac{ig_{s}G_{ab}^{\theta\eta}}{32\pi^{2}x^{2}}\Big[x\!\!\!/\sigma_{\theta\eta} +\sigma_{\theta\eta}x\!\!\!/ \Big]
 \nonumber\\&-&\frac{x\!\!\!/ x^{2}g_s^2}{7776}\langle
 \overline{q}q\rangle^2\delta_{ab}-\frac{x^4\langle
 \overline{q}q\rangle\langle
 g_s^2G^2\rangle}{27648}\delta_{ab}+\frac{m_q}{32\pi^2}[ln(\frac{-x^2\Lambda^2}{4})+2 \gamma_E]g_{s}G_{ab}^{\theta\eta}\sigma_{\theta\eta}+\cdots,
\end{eqnarray}
in $x$ space with the Euler constant, $\gamma_E \simeq 0.577$. The
parameter $\Lambda$ is a scale parameter separating the
perturbative and nonperturbative regions. After the insertion of
the propagator into Eq.~(\ref{corre4}) and performing the Fourier
and Borel transformations as well as continuum subtraction, for
the QCD side of the correlation function, corresponding to the
coefficients of the selected structures, we get

\begin{eqnarray}
\Pi_{1(+)}^B=-\Pi_{1(-)}^B&= &\frac{1}{\pi^{2}}\int_0^{s_{0}} ds e^{-\frac{s}{M^2}}
\left\lbrace \frac{s^2}{5\times 2^{5}\pi^{2}}- \frac{3\langle
\bar{s} s\rangle m_s}{ 2^{2}} -\frac{5 \langle g_s^2
G^2\rangle}{3^2\times 2^6 \pi^2} +\frac{ \langle g_s^2
G^2\rangle\langle \bar{s} s\rangle m_s}{3^2 M^4}
\mathrm{Log}\left[\frac{s}{\Lambda^{2}} \right] \right\rbrace
\nonumber \\
&+&  \frac{3m_0^{2}\langle \bar{s} s\rangle m_s}{ 2^3
\pi^2}+\frac{4\langle \bar{s} s\rangle^2}{3}  -\frac{7 m_0^2
\langle \bar{s} s\rangle^2}{3^2 M^2} +\frac{ \langle g_s^2
G^2\rangle\langle \bar{s} s\rangle m_s}{3\times 2^3 \pi^2 M^2}
  -\frac{53\langle g_s^2 G^2\rangle m_0^2 \langle \bar{s} s\rangle m_s}{3^3\times
   2^6 \pi^2 M^4}
   \nonumber \\
&+& \frac{ \langle g_s^2 G^2\rangle \langle \bar{s} s\rangle
m_s}{3^2 \pi^2 s_0 M^2} \left[M^2+s_0
\mathrm{Log}\left[\frac{s}{\Lambda^{2}} \right]\right]
e^{-\frac{s_0}{M^2}}, \label{Coefqslashgmunu}
\end{eqnarray}
and
\begin{eqnarray}
\Pi_{2(+)}^B=\Pi_{2(-)}^B&=&\frac{1}{\pi^{2}}\int_0^{s_{0}} ds e^{-\frac{s}{M^2}}
\left\lbrace \frac{3 s^2 m_s  }{2^{6} \pi^{2}} - \frac{s \langle
\bar{s} s\rangle }{3}+\frac{m_{0}^{2} \langle \bar{s} s\rangle
}{2\times3} + \frac{5\langle g_{s}^{2}G^{2}\rangle m_s }{3\times
2^7 \pi^2}\left[ \left(2\gamma_{E}-1 \right)-2
\mathrm{Log}\left[\frac{s}{\Lambda^{2}} \right]  \right] \right.
\nonumber \\
&+&\left. \frac{\langle g_{s}^{2}G^{2}\rangle^{2} m_s }{3^2\times
2^{8} \pi^2 M^4}\mathrm{Log}\left[\frac{s}{\Lambda^{2}}\right]
 \right\rbrace-\frac{5\gamma_E \langle g_{s}^{2}G^{2}\rangle M^2 m_s }
 {3\times2^6 \pi^4}
 + 2\langle \bar{s} s\rangle^2 m_s+\frac{\langle
\bar{s} s\rangle\langle g_{s}^{2}G^{2}\rangle}{3^2\times
2^2\pi^2}+\frac{5\langle g_{s}^{2}G^{2}\rangle^2 m_s}{3^3\times2^8
\pi^2M^2}
\nonumber \\
&-& \frac{11m_0^2\langle \bar{s} s\rangle^2 m_s}{M^2}-\frac{m_0^2
\langle \bar{s} s\rangle\langle g_{s}^{2}G^{2}\rangle}{3^2\times
2^5 \pi^2 M^2}+\frac{\langle \bar{s} s\rangle^2\langle
g_{s}^{2}G^{2}\rangle m_s}{3\times2^2 M^4}+\frac{m_0^2\langle
\bar{s} s\rangle^2\langle g_{s}^{2}G^{2}\rangle m_s}{3\times 2^2
M^6}
\nonumber \\
&+&  \left[\frac{5\gamma_E \langle g_{s}^{2}G^{2}\rangle M^2
m_s}{3\times2^6\pi^4}+\frac{\langle g_{s}^{2}G^{2}\rangle^2
m_s}{3^2\times2^8 M^2
s_0}\left(M^2+s_0\mathrm{Log}\left[\frac{s_0}{\Lambda^{2}}\right]
\right) \right]e^{-\frac{s_0}{M^2}}, \label{Coefgmunu}
\end{eqnarray}

At this stage the calculations of the physical and QCD sides are completed and we need to match the coefficients of the same structures obtained from both sides which give us the following sum rules:
\begin{eqnarray}
\lambda_{+}^2 e^{-\frac{m_{+}^{2}}{M^{2}}}+\lambda_{-}^2 e^{-\frac{m_{-}^{2}}{M^{2}}}&=&{\Pi}_{1(+)}^{B},
\nonumber \\
m_{+} \lambda_{+}^2 e^{-\frac{m_{+}^{2}}{M^{2}}}-m_{-} \lambda_{-}^2 e^{-\frac{m_{-}^{2}}{M^{2}}}&=&{\Pi}_{2(+)}^{B}.
\label{Eq:sumrule}
\end{eqnarray}
\begin{eqnarray}
\lambda_{+}^{\prime 2} e^{-\frac{m_{+}^{2}}{M^{2}}}+\lambda_{-}^{\prime 2} e^{-\frac{m_{-}^{2}}{M^{2}}}&=&{\Pi}_{1(-)}^{B},
\nonumber \\
-m_{+} \lambda_{+}^{\prime 2} e^{-\frac{m_{+}^{2}}{M^{2}}}+m_{-} \lambda_{-}^{\prime 2} e^{-\frac{m_{-}^{2}}{M^{2}}}&=&{\Pi}_{2(-)}^{B}.
\label{Eq:sumrule2}
\end{eqnarray}
In Eq.~(\ref{Eq:sumrule}) there are three unknown parameters which are the mass and residue of the   orbitally excited state $\Omega$ as well as the residue of the ground state $\Omega$. For determination of these three parameters we need at least three equations. Therefore, the third equation  is obtained from the first one given in Eq.~(\ref{Eq:sumrule}) by performing derivative with respect to $(-\frac{1}{M^2})$ variable. After some calculations for the mass and the residue of excited state $\Omega$ from Eq.~(\ref{Eq:sumrule}) we get
%In present calculation we take the mass of the ground state $\Omega$ as input and obtain the three %unknown parameters. One of them is ground state residue $\lambda_{\Omega}$ of $\Omega$ and the other two %are residue $\lambda_{\widetilde{\Omega}}$ and mass $m_{\widetilde{\Omega}}$ of orbitally excited state %$\widetilde{\Omega}$. To get them  we solve the expressions given in Eq.~(\ref{Eq:sumrule}) together %with their derivatives with respect to $1/M^2$, since they are not enough to get three unknowns. The %residue of the ground state $\Omega$ was already calculated in Ref.~\cite{Aliev:2016jnp} therefore we %ill not concentrate on it. We get the following expressions for the residue $%\lambda_{\widetilde{\Omega}}$ and mass $m_{\widetilde{\Omega}}$
%
\begin{eqnarray}
 m_{-}^2&=&\frac{\frac{d}{d(-\frac{1}{M^2})}({\Pi}_{2(+)}^{B}-m_{+}{\Pi}_{1(+)}^{B})}{{\Pi}_{2(+)}^{B}-m_{+}{\Pi}_{1(+)}^{B}}\nonumber\\
 \lambda_{-}^2&=&\frac{{\Pi}_{2(+)}^{B}-m_{+}{\Pi}_{1(+)}^{B}}{m_{+}+m_{-}}e^{\frac{m_{-}}{M^2}}.
 \label{Eq:massresidue}
\end{eqnarray}
Together with the mass of the ground state $\Omega$, some of the input parameters that we need to perform the numerical analysis are given in Table~\ref{tab:Param}.
\begin{table}[tbp]
%\rowcolors{1}{lightgray}{white}
\begin{tabular}{|c|c|}
\hline\hline
Parameters & Values \\ \hline\hline
$m_{\Omega}$                             & $1672.45\pm 0.29~\mathrm{MeV}$ \cite{Patrignani:2016xqp}\\
$m_{s}$                                  & $128^{+12}_{-4}~\mathrm{MeV}$ \cite{Patrignani:2016xqp}\\
$\langle \bar{q}q \rangle (1\mbox{GeV})$ & $(-0.24\pm 0.01)^3$ $\mathrm{GeV}^3$ \cite{Belyaev:1982sa}  \\
$\langle \bar{s}s \rangle $              & $0.8\langle \bar{q}q \rangle$ \cite{Belyaev:1982sa} \\
$m_{0}^2 $                               & $(0.8\pm0.1)$ $\mathrm{GeV}^2$ \cite{Belyaev:1982sa}\\
$\langle g_s^2 G^2 \rangle $             & $4\pi^2 (0.012\pm0.004)$ $~\mathrm{GeV}
^4 $\cite{Belyaev:1982cd}\\
$ \Lambda $                              & $ (0.5\pm0.1) $ $\mathrm{GeV} $ \cite{Chetyrkin:2007vm} \\
\hline\hline
\end{tabular}%
\caption{Some input parameters.}
\label{tab:Param}
\end{table}
Note that in the Table~\ref{tab:Param} the mass of the $s$ quark is presented rescaling it to the normalization point $\mu_0^2=1~\mbox{GeV}^2$. In addition to these parameters, sum rules contain two auxiliary parameters, Borel mass parameter $M^2$ and continuum threshold $s_0$. The physical quantities should be practically independent of these parameters. To obtain a working window for $M^2$, we require the pole dominance over the contributions of higher states and continuum. And also the results coming from higher dimensional operators should contribute less than the lower dimensional ones, since OPE should be convergent. These requirements lead to the following working window for the Borel parameter
\begin{eqnarray}
3.0~\mbox{GeV}^2\leq M^2\leq 4.0~\mbox{GeV}^2.
\end{eqnarray}
For the threshold parameter we choose the  interval
\begin{eqnarray}
7.3~\mbox{GeV}^2\leq s_0\leq 8.4~\mbox{GeV}^2,
\end{eqnarray}
which leads to  a relatively weak dependence of the  results on the threshold parameter.

\begin{figure}[h!]
\begin{center}
\includegraphics[totalheight=5cm,width=7cm]{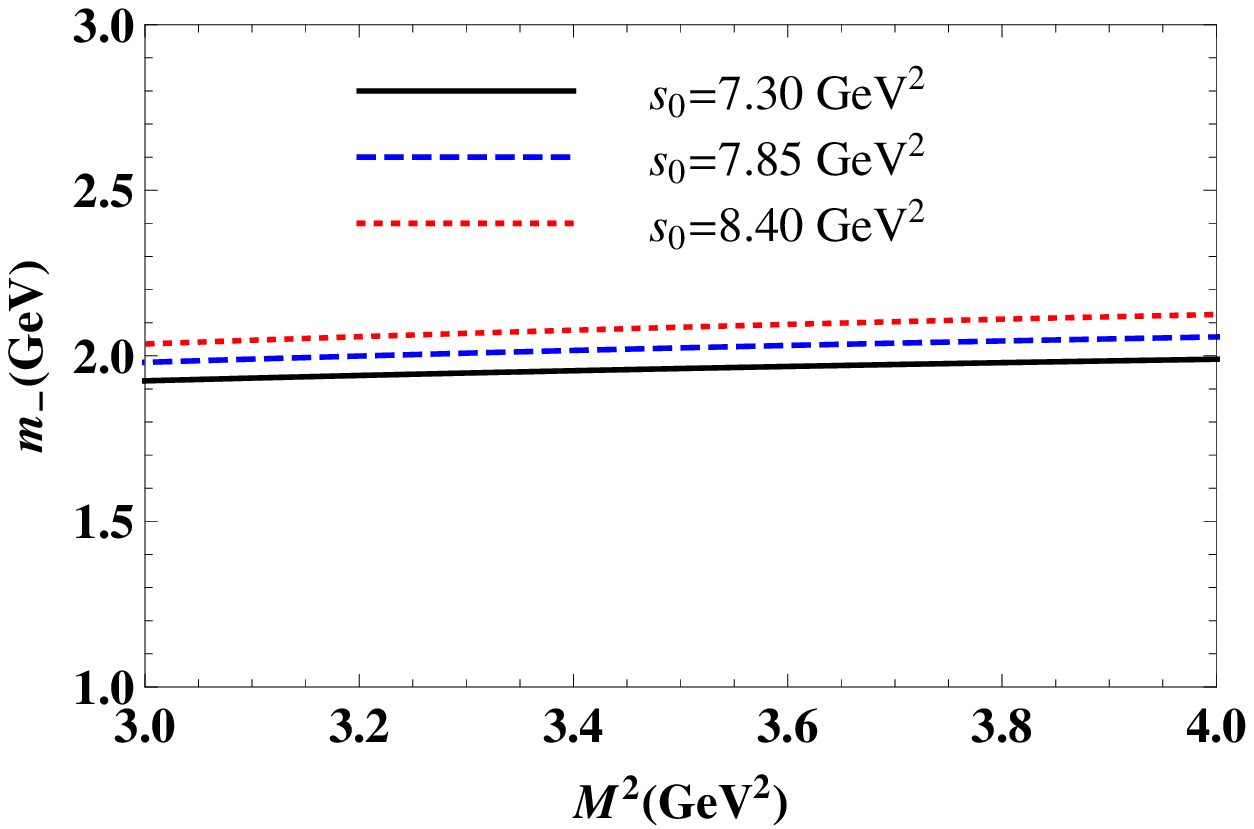}
\includegraphics[totalheight=5cm,width=7cm]{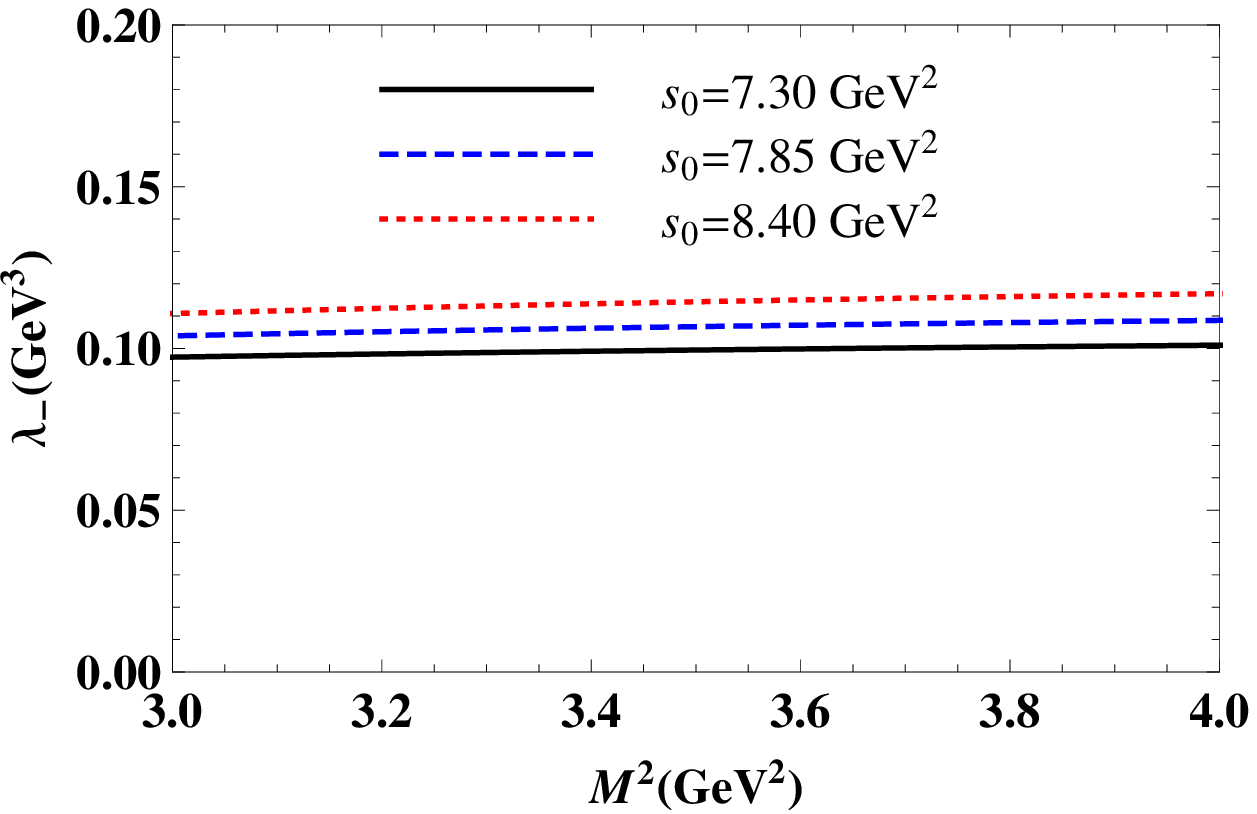}
\end{center}
\caption{\textbf{Left:} The mass of the orbitally excited $\Omega$  baryon vs Borel
parameter $M^2$.
\textbf{Right:} The residue  of the orbitally excited $\Omega$   baryon vs Borel
parameter $M^2$. }
\label{gr:OPEvsMsq}
\end{figure}

\begin{figure}[h!]
\begin{center}
\includegraphics[totalheight=5cm,width=7cm]{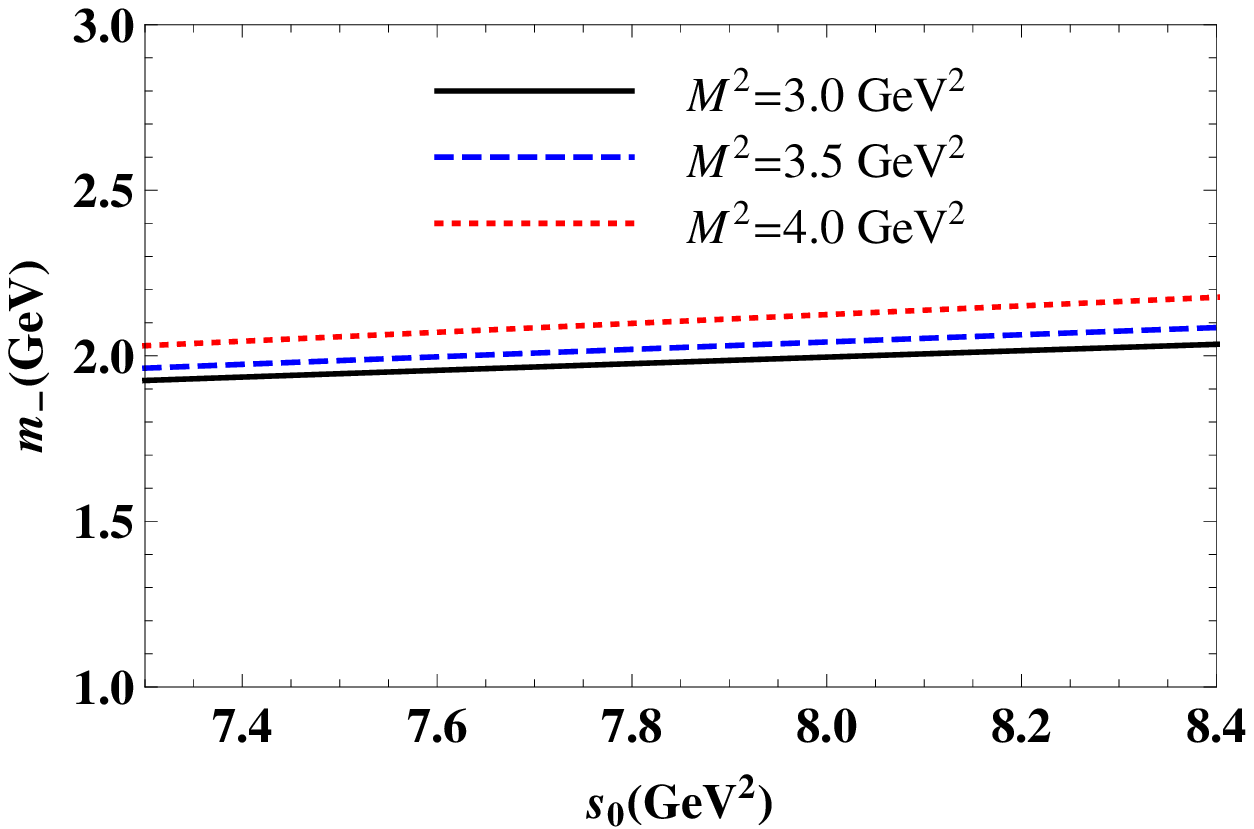}
\includegraphics[totalheight=5cm,width=7cm]{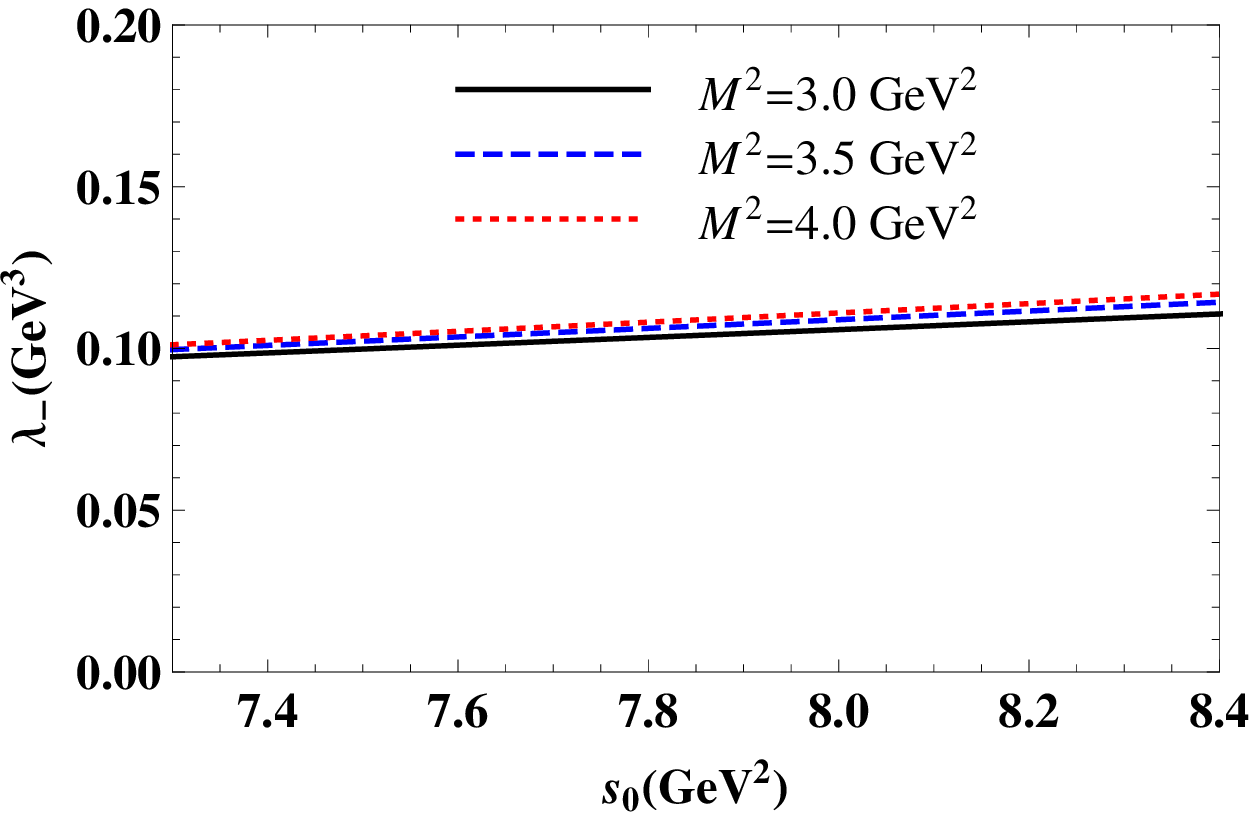}
\end{center}
\caption{\textbf{Left:} The mass of the orbitally excited $\Omega$  baryon vs threshold parameter $s_0$.
\textbf{Right:} The residue  of the orbitally excited $\Omega$   baryon vs threshold
parameter $s_0$. }
\label{gr:OPEvss0}
\end{figure}

Using the above working windows of auxiliary  parameters we show the dependencies of the mass and residue  of the  $\Omega(2012)$ state given in Eq.~(\ref{Eq:massresidue}) as  a function of $M^2$ at fixed values of $s_0$ and  as a function of $s_0$ at fixed values of $M^2$ in Figs.~\ref{gr:OPEvsMsq} and \ref{gr:OPEvss0}, respectively.  We observe that the dependencies of the mass and residue of the excited state of $\Omega$ on the auxiliary parameters are rather weak in their working intervals.

Using the positive parity current, we get our final results of mass and residue for excited $\Omega$ state as
\begin{eqnarray}
m_{-}=2019^{+17}_{-29}~\mbox{MeV} ~~~~~~~~~~~\lambda_{-}=0.108^ {+0.004}_{-0.005}~\mbox{GeV}^3.
\end{eqnarray}
We perform similar analysis for the mass and residue of excited $\Omega$ state using the negative parity current and Eq.~(\ref{Eq:sumrule2}). Our final predictions in this case are 
\begin{eqnarray}
m_{-}=2020^{+19}_{-28}~\mbox{MeV} ~~~~~~~~~~~\lambda^{\prime}_{-}=0.094^ {+0.003}_{-0.004}~\mbox{GeV}^3.
\end{eqnarray}
The errors in the results are due to the uncertainties carried by the input parameters as well as those coming from the working windows for auxiliary parameters. As is seen, the obtained results for the mass predicted from the positive and negative parity currents are in a nice consistency with the experimental value, $2012.4 \pm 0.7$~(stat)$\pm 0.6$~(syst)~MeV measured by the Belle Collaboration.

In summary, inspired by the recent discovery of the Belle Collaboration we calculated the mass and residue of the  $\Omega(2012)$ state by using two different forms of interpolating current within the QCD sum rule approach. We found that the mass prediction is insensitive to the choice of the interpolating current. We compared the obtained result for the mass of this state with the experimental value, which allowed us to assign the quantum numbers $J^P=\frac{3}{2}^-$ for the $\Omega(2012)$ state. The result obtained for the residue of this state can be used in determinations of its electromagnetic properties as well as many parameters related to different decays of this particle.

\section*{ACKNOWLEDGEMENTS}

H. S. thanks Kocaeli University for the partial financial support through the grant BAP 2018/070.
K. A. appreciates  financial support of  Dogus University  through the grant BAP 2015-16-D1-B04.

\label{sec:Num}
%%%%%%%%%%%%%%%%%%%%%%%%%%%%%%%%%%%%%%%%%%%%%%%%%%%%%%%%%%%%%%%%%%%%%%%%%%%%%%%%%%%%%%%%%%%%%%%%%%%%%%%%%%%%


\begin{thebibliography}{99}
%\bibitem{} %\cite{FirstTheor}

\bibitem{Yelton:2018mag}
  J.~Yelton {\it et al.} [Belle Collaboration],
  Observation of an excited $\Omega^-$ baryon,
  %Submitted to: Phys.Rev.Lett.
  [arXiv:1805.09384 [hep-ex]].

  \bibitem{Patrignani:2016xqp}
  C.~Patrignani {\it et al.} [Particle Data Group],
  Review of Particle Physics,
  Chin.\ Phys.\ C {\bf 40}, 100001 (2016).
  %doi:10.1088/1674-1137/40/10/100001


\bibitem{Chao:1980em}
  K.~T.~Chao, N.~Isgur and G.~Karl,
  Strangeness -2 and -3 Baryons in a Quark Model With Chromodynamics,
  Phys.\ Rev.\ D {\bf 23}, 155 (1981).
  %doi:10.1103/PhysRevD.23.155
  %%CITATION = doi:10.1103/PhysRevD.23.155;%%


\bibitem{Kalman:1982ut}
  C.~S.~Kalman,
  $P$ Wave Baryons in a Consistent Quark Model With Hyperfine Interactions,
  Phys.\ Rev.\ D {\bf 26}, 2326 (1982).
  %doi:10.1103/PhysRevD.26.2326
  %%CITATION = doi:10.1103/PhysRevD.26.2326;%%
  %15 citations counted in INSPIRE as of 29 May 2018

  \bibitem{Capstick:1986bm}
  S.~Capstick and N.~Isgur,
  Baryons in a Relativized Quark Model with Chromodynamics,
  Phys.\ Rev.\ D {\bf 34}, 2809 (1986).
  %[AIP Conf.\ Proc.\  {\bf 132}, 267 (1985)].
  %doi:10.1103/PhysRevD.34.2809, 10.1063/1.35361
  %%CITATION = doi:10.1103/PhysRevD.34.2809, 10.1063/1.35361;%%
  %1121 citations counted in INSPIRE as of 26 May 2018


\bibitem{Loring:2001ky}
  U.~Loring, B.~C.~Metsch and H.~R.~Petry,
  The Light baryon spectrum in a relativistic quark model with instanton induced quark forces: The Strange baryon spectrum,
  Eur.\ Phys.\ J.\ A {\bf 10}, 447 (2001)
  %doi:10.1007/s100500170106
  [hep-ph/0103290].
  %%CITATION = doi:10.1007/s100500170106;%%
  %143 citations counted in INSPIRE as of 26 May 2018


\bibitem{Liu:2007yi}
  J.~Liu, R.~D.~McKeown and M.~J.~Ramsey-Musolf,
  Global Analysis of Nucleon Strange Form Factors at Low $Q^2$,
  Phys.\ Rev.\ C {\bf 76}, 025202 (2007)
  %doi:10.1103/PhysRevC.76.025202
  [arXiv:0706.0226 [nucl-ex]].
  %%CITATION = doi:10.1103/PhysRevC.76.025202;%%
  %65 citations counted in INSPIRE as of 26 May 2018


\bibitem{Pervin:2007wa}
  M.~Pervin and W.~Roberts,
  Strangeness -2 and -3 baryons in a constituent quark model,
  Phys.\ Rev.\ C {\bf 77}, 025202 (2008).
  %doi:10.1103/PhysRevC.77.025202
  [arXiv:0709.4000 [nucl-th]].
  %%CITATION = doi:10.1103/PhysRevC.77.025202;%%
  %23 citations counted in INSPIRE as of 29 May 2018




\bibitem{An:2013zoa}
  C.~S.~An, B.~C.~Metsch and B.~S.~Zou,
  Mixing of the low-lying three- and five-quark $\Omega$ states with negative parity,
  Phys.\ Rev.\ C {\bf 87}, 065207 (2013)
 % doi:10.1103/PhysRevC.87.065207
  [arXiv:1304.6046 [hep-ph]].

\bibitem{An:2014lga}
  C.~S.~An and B.~S.~Zou,
  Low-lying $\Omega$ states with negative parity in an extended quark model with Nambu-Jona-Lasinio interaction,
  Phys.\ Rev.\ C {\bf 89}, 055209 (2014)
 % doi:10.1103/PhysRevC.89.055209
  [arXiv:1403.7897 [hep-ph]].
  %%CITATION = doi:10.1103/PhysRevC.89.055209;%%

\bibitem{Faustov:2015eba}
  R.~N.~Faustov and V.~O.~Galkin,
  Strange baryon spectroscopy in the relativistic quark model,
  Phys.\ Rev.\ D {\bf 92}, 054005(2015)
  %doi:10.1103/PhysRevD.92.054005
  [arXiv:1507.04530 [hep-ph]].

  \bibitem{Engel:2013ig}
  G.~P.~Engel {\it et al.} [BGR Collaboration],
  QCD with Two Light Dynamical Chirally Improved Quarks: Baryons,
  Phys.\ Rev.\ D {\bf 87}, 074504 (2013)
  %doi:10.1103/PhysRevD.87.074504
  [arXiv:1301.4318 [hep-lat]].
  %%CITATION = doi:10.1103/PhysRevD.87.074504;%%
  %70 citations counted in INSPIRE as of 26 May 2018

%\cite{Liang:2015bxr}
\bibitem{Liang:2015bxr}
  J.~Liang {\it et al.} [CLQCD Collaboration],
  Spectrum and Bethe-Salpeter amplitudes of $\Omega$ baryons from lattice QCD,
  Chin.\ Phys.\ C {\bf 40}, 041001 (2016)
 % doi:10.1088/1674-1137/40/4/041001
  [arXiv:1511.04294 [hep-lat]].

  \bibitem{Oh:2007cr}
  Y.~Oh,
  $\Xi$ and $\Omega$ baryons in the Skyrme model,
  Phys.\ Rev.\ D {\bf 75}, 074002 (2007)
  %doi:10.1103/PhysRevD.75.074002
  [hep-ph/0702126 [HEP-PH]].

  \bibitem{Xiao:2018pwe}
  L.~Y.~Xiao and X.~H.~Zhong,
  A possible interpretation of the newly observed $\Omega(2012)$ state,
  arXiv:1805.11285 [hep-ph].

  \bibitem{Kaxiras:1985zv}
  E.~Kaxiras, E.~J.~Moniz and M.~Soyeur,
  Hyperon Radiative Decay,
  Phys.\ Rev.\ D {\bf 32}, 695 (1985).

  \bibitem{Aliev:2014pfa}
  T.~M.~Aliev and M.~Savci,
  Magnetic moments of negative-parity baryons in QCD,
  Phys.\ Rev.\ D {\bf 89}, 053003 (2014)
  %doi:10.1103/PhysRevD.89.053003
  [arXiv:1402.4609 [hep-ph]].

  \bibitem{Aliev:2015}
  T.~M.~Aliev and M.~Savci,
  Magnetic moments of $J^P=\frac{3}{2}^-$ baryons in QCD
  Phys.\ Rev.\ D {\bf 90}, 116006 (2015).


\bibitem{Aliev:2016jnp}
  T.~M.~Aliev, K.~Azizi and H.~Sundu,
  Radial Excitations of the Decuplet Baryons,
  Eur.\ Phys.\ J.\ C {\bf 77}, no. 4, 222 (2017)
 % doi:10.1140/epjc/s10052-017-4782-0
  [arXiv:1612.03661 [hep-ph]].

  %\cite{Shifman:1978bx}
\bibitem{Shifman:1978bx}
  M.~A.~Shifman, A.~I.~Vainshtein and V.~I.~Zakharov,
  QCD and Resonance Physics. Theoretical Foundations,
  Nucl.\ Phys.\ B {\bf 147}, 385 (1979).
  %\cite{Shifman:1978by}

\bibitem{Shifman:1978by}
  M.~A.~Shifman, A.~I.~Vainshtein and V.~I.~Zakharov,
  QCD and Resonance Physics: Applications,
  Nucl.\ Phys.\ B {\bf 147}, 448 (1979).

\bibitem{Ioffe81} B.~L.~Ioffe, Calculation of Baryon Masses in Quantum
Chromodynamics, Nucl. Phys. B {\bf 188}, 317 (1981) Erra-
tum: [Nucl. Phys. B {\bf 191}, 591 (1981)].


\bibitem{Belyaev:1982sa}
  V.~M.~Belyaev and B.~L.~Ioffe,
  Determination of Baryon and Baryonic Resonance Masses from QCD Sum Rules. 1. Nonstrange Baryons,
  Sov.\ Phys.\ JETP {\bf 56}, 493 (1982)
  [Zh.\ Eksp.\ Teor.\ Fiz.\  {\bf 83}, 876 (1982)].
  %%CITATION = SPHJA,56,493;%%
  %347 citations counted in INSPIRE as of 05 Dec 2016

\bibitem{Belyaev:1982cd}
  V.~M.~Belyaev and B.~L.~Ioffe,
  Determination of the baryon mass and baryon resonances from the quantum-chromodynamics sum rule. Strange baryons,
  Sov.\ Phys.\ JETP {\bf 57}, 716 (1983)
  [Zh.\ Eksp.\ Teor.\ Fiz.\  {\bf 84}, 1236 (1983)].

%\cite{Chetyrkin:2007vm}
\bibitem{Chetyrkin:2007vm}
  K.~G.~Chetyrkin, A.~Khodjamirian and A.~A.~Pivovarov,
  Towards NNLO Accuracy in the QCD Sum Rule for the Kaon Distribution Amplitude,
  Phys.\ Lett.\ B {\bf 661} (2008) 250
%  doi:10.1016/j.physletb.2008.02.031
  [arXiv:0712.2999 [hep-ph]].



\end{thebibliography}
\end{document}